\documentclass[paper]{JHEP3}

\usepackage{epsfig,multicol,multirow}
\usepackage{amsmath,subfigure,latexsym,amssymb}

\def\lsi{\raise0.3ex\hbox{$<$\kern-0.75em\raise-1.1ex\hbox{$\sim$}}}
\def\gsi{\raise0.3ex\hbox{$>$\kern-0.75em\raise-1.1ex\hbox{$\sim$}}}

\newcommand\fverb{\setbox\pippobox=\hbox\bgroup\verb}
\newcommand\fverbdo{\egroup\medskip\noindent%
                        \fbox{\unhbox\pippobox}\ }
\newcommand\fverbit{\egroup\item[\fbox{\unhbox\pippobox}]}
\newbox\pippobox
\newcommand{\beq}{\begin{equation}}
\newcommand{\eeq}{\end{equation}}
\newcommand{\beqa}{\begin{eqnarray}}
\newcommand{\eeqa}{\end{eqnarray}}
\newcommand{\n}{\nonumber \\}

\newcommand{\Z}{\mathbb{Z}}

\newcommand{\id}{{1\!\!1}} 
\def\e{{\,\rm e}\,}

\newcommand{\del}{\partial}
\newcommand {\tr}{{\rm tr\,}}
\newcommand {\Tr}{\mbox{Tr\,}}
\newcommand {\CTr}{{\cal T} \! r\,}

\preprint{
SAGA-HE-247 \\ 
KEK-TH-1279
}
\title{
Finite-matrix
formulation of gauge theories\\
on a non-commutative torus\\
with twisted boundary conditions
}

\author{
Hajime Aoki${}^{a}$,
Jun Nishimura${}^{bc}$ and
Yoshiaki Susaki${}^{b}$\\ 
\llap{$^a$}Department of Physics, Saga University, 
Saga 840-8502, Japan \\
\llap{$^b$}High Energy Accelerator Research Organization (KEK), \\
Tsukuba, Ibaraki, 305-0801, Japan \\
\llap{$^c$}Department of Particle and Nuclear Physics,\\
Graduate University for Advanced Studies (SOKENDAI),\\
Tsukuba, Ibaraki 305-0801, Japan \\
\email{haoki@cc.saga-u.ac.jp,
jnishi@post.kek.jp,
susaki@post.kek.jp}} 

\abstract{
We present a novel finite-matrix formulation of 
gauge theories on a non-commutative torus.
Unlike the previous formulation based on
a map from a square matrix to a field on a discretized torus
with periodic boundary conditions,
our formulation is based on the algebraic 
characterization of the configuration space.
This enables us to
describe the twisted boundary conditions
in terms of finite matrices
and 
hence to realize the Morita equivalence
at a fully regularized level.
Matter fields in the fundamental representation
turn out to be represented by \emph{rectangular} matrices
for twisted boundary conditions
analogously to the matrix spherical harmonics
on the fuzzy sphere with the monopole background.
The corresponding Ginsparg-Wilson Dirac operator
defines an index, which can be used to
classify gauge field configurations
into topological sectors.
We also perform Monte Carlo calculations
for the index
as a consistency check.
Our formulation is expected to be useful for
applications of non-commutative geometry
to various problems related to topological aspects
of field theories and string theories.
}


\begin{document}

\section{Introduction}

Non-commutative (NC) geometry \cite{Sny,Connes}
was proposed long time ago
as a simple modification of our notion of space-time
at short distances possibly due to effects of 
quantum gravity \cite{gravity}.
It has attracted much attention since 
field theories on a NC geometry
were shown to appear naturally from matrix models \cite{CDS,AIIKKT}
and from string theories \cite{String}.
In fact it turned out that the NC geometry
affects not only the short-distance physics
but also the long-distance physics through loop effects.
This UV/IR mixing phenomenon \cite{rf:MRS}
makes such field theories
more interesting due to rich physics,
but it also poses both technical and conceptual problems
in various applications.
In particular,
the appearance of a new type of IR divergence
spoils the perturbative renormalizability in most cases 
\cite{Chepelev:1999tt}.
Therefore, nonperturbative studies of NC field theories
based on a suitable regularization are highly motivated.

Not surprisingly, considering the links to string theory,
a natural regularization of NC field theories
is possible by representing fields by finite matrices.
In the case of a NC torus, for instance, 
the so-called twisted Eguchi-Kawai model \cite{EK,GAO}
is reinterpreted
as a lattice formulation of NC field theories \cite{AMNS},
in which a finite $N$ matrix is mapped
one-to-one onto a field on a discretized torus.
Monte Carlo simulations based on this formulation 
have been performed intensively.
For instance, in ref.\ \cite{2dU1} the existence of 
a sensible continuum limit (and hence the
nonperturbative renormalizability) 
of a NC field theory was shown for the first time
in the case of U(1) gauge theory on a 2d NC torus.
Spontaneous breaking of the translational symmetry 
\cite{GuSo,rf:MVR,rf:AL},
which occurs as a result of the UV/IR mixing effects,
has been studied nonperturbatively
in NC scalar field theories \cite{AC,Bietenholz:2004xs}
and 
in NC U(1) gauge theory in 4d \cite{4dU1}.
This phenomenon may be viewed as the collapsing of the
NC manifolds \cite{Azuma:2004zq,Azeyanagi:2008bk}. 

%
%

Another interesting feature of NC field theories
is the appearance of a new type of topological objects
\cite{NCsoliton}.
We consider it important to study  
dynamical properties of these objects,
since they may play an important role
in physical problems
such as the strong CP problem and the
baryon number non-conservation in the electroweak theory,
which are expected to be related to topological aspects 
of field theories.
They may also be relevant to string theory, in which
chiral fermions on our space-time 
are considered to be realized 
by compactification with a nontrivial index
in the extra dimensions. 
In order to address such dynamical issues,
we need to perform nonperturbative studies based on 
a regularized formulation.
It turned out, however, that finite-matrix description of
topologically nontrivial configurations 
is not that straightforward.
For instance, refs.\ \cite{non-trivial_config,Balachandran:2003ay} use
mathematical devices such as
the projective modules and the algebraic K-theory.

The difficulty is related to the fact that
topologically nontrivial configurations
in the \emph{commutative} space typically allow two descriptions;
one in terms of single-valued functions with singularities,
and the other in terms of multi-valued functions without
singularities. These two descriptions are related to
each other through a singular gauge transformation.
In the case of NC geometry the first description is somehow 
prohibited due to smearing effects.
This is demonstrated on a NC torus
with periodic boundary conditions
\cite{Aoki:2006sb,Aoki:2006zi,Frisch:2007zz},
where topologically nontrivial configurations are shown to be
suppressed in the continuum limit.\footnote{Similar results 
were obtained in the fuzzy sphere case \cite{Aoki:2008qta}.}
The situation is in sharp contrast to that in the 
commutative space as seen in lattice simulations \cite{GHL}.
Thus we are led to consider the second description.

As in the commutative case,
one can
think of 
twisting boundary conditions \cite{'tHooft:1979uj},
which corresponds to introducing a background magnetic flux.
However, the finite-matrix formulation of gauge theories
on a NC torus proposed in ref.\ \cite{AMNS}
essentially deals with
the case of periodic boundary conditions.
In this paper we reconsider the formulation
from a purely algebraic point of view,\footnote{See
ref.\ \cite{Griguolo:2003kq} for a 
reformulation with different motivations.
}
and generalize it in such
a way that it
allows for explicit description of 
twisted boundary conditions in terms of finite matrices.
%
%
%
%
%
%
Our new formulation
realizes the Morita equivalence at a fully regularized level.
%
Matter fields in 
the fundamental representation can be naturally described
by finite rectangular matrices in the presence of the
magnetic flux.
This is analogous to the situation 
in the fuzzy sphere case with
the monopole background,
where rectangular matrices appear as 
the matrix version of the monopole harmonics
\cite{Balachandran:2003ay,Aoki:2006wv,Ishiki:2006yr}.
The corresponding Ginsparg-Wilson Dirac operator enables
us to define an index, which can be used to classify regularized 
gauge field configurations into topological sectors.
We present Monte Carlo results for the index, which demonstrate
that topologically nontrivial configurations indeed 
survive the continuum limit in a 2d gauge theory
with twisted boundary conditions.

In fact there is yet another possibility 
for realizing topologically
nontrivial configurations, 
in which they are described by 
both single-valued and regular functions.
The idea is to 
consider the Higgs phase of gauge theories, 
where the winding number of the Higgs field 
substitutes the role
of the twisted boundary conditions.
While we do not discuss this possibility in this paper,
let us comment on some recent developments
in this direction.
In the case of fuzzy sphere,
the corresponding finite-matrix formulation was constructed,
and 
the index of the Ginsparg-Wilson Dirac operator
involving the Higgs field,
which reproduces
the Higgs winding number,
was formulated for general background configurations
\cite{Aoki:2006wv,Aoki:2008qta}.
An explicit NC configuration corresponding to the 
't Hooft-Polyakov monopole\footnote{Dynamical 
properties of these configurations were studied \cite{Azuma:2004zq}
and their instability was interpreted as
a dynamical generation of nontrivial topological sectors,
which may be used 
for realizing chiral fermions on our space-time \cite{AIMN}.
}
was constructed \cite{Balachandran:2003ay,AIN3},
and the spectrum of matter fluctuations 
around this background was obtained
\cite{Balachandran:2003ay,Aoki:2006wv}.
The matrix version of monopole harmonics,
which appears here,
plays an important role also in 
a nonperturbative formulation of super Yang-Mills theory
on $R \times S^3$, where
$S^3$ is regarded as an $S^1$ bundle 
over $S^2$ \cite{Ishii:2008ib}.
%
We expect that the ideas to use regular configurations
for describing nontrivial topological sectors in NC geometry
by considering the winding Higgs field or the 
twisted boundary conditions,
provide physical understanding
to the previous mathematical formulations 
\cite{non-trivial_config,Balachandran:2003ay} mentioned above.

The rest of this paper is organized as follows.
In section \ref{sec:morita} 
we briefly review the gauge theories on a NC torus
with twisted boundary conditions.
A more detailed review is 
provided in appendix \ref{section:review}.
In section \ref{sec:solvetbc} we rewrite the
boundary conditions in a covariant form.
In section \ref{sec:mmformulation}
we characterize the configuration
space of a regularized field 
as a representation space of the algebra
of the coordinate and shift operators.
In section \ref{sec:constcoodshiftope}
we explicitly construct the representation space of the algebra
to arrive at a finite-matrix formulation for twisted 
boundary conditions.
In section \ref{sec:actiongauge} we describe
the actions for the gauge field and for
fundamental matters.
In section \ref{sec:numerical} we present 
Monte Carlo results for the index of the Ginsparg-Wilson
Dirac operator. 
Section \ref{sec:summary} is devoted
to a summary and discussions.

\section{Brief review of gauge theories on a NC torus}
\label{sec:morita}

In this section we briefly review 
gauge theories on a continuous NC torus with twisted boundary conditions,
which form the basis of our finite-matrix formulation.
The readers who are not familiar with the subject
are recommended to read appendix \ref{section:review}, 
where we provide a more detailed and self-contained review.
There one can also find some derivations omitted in this section,
descriptions in terms of fields instead of operators,
and an explicit form of the actions.
%

In NC geometry the coordinate operators $\hat x_\mu$ and
the derivative operators $\hat\partial_\mu$ 
satisfy the algebra
\beq
\left[\hat x_\mu \,,\,\hat x_\nu\right]=i\,\theta_{\mu\nu}~,~~
\left[\hat\partial_\mu\,,\,\hat x_\nu\right]=\delta_{\mu\nu}~,~~
\label{noncommalg}
\eeq
where $\theta_{\mu\nu}$ represents the 
non-commutativity of the space-time.
Let us consider a $D$-dimensional torus
with the period $L$.
When we impose periodic boundary conditions,
we consider the operators 
$\hat Z_\mu = e^{2\pi i \hat x_\mu/L}$
instead of $\hat x_\mu$,
and the algebra (\ref{noncommalg}) becomes
\beqa
\hat Z_\mu \hat Z_\nu &=& e^{-2\pi i \Theta_{\mu\nu}}
\hat Z_\nu \hat Z_\mu \ ,
\label{ncalgZZ}\\
\left[\hat\partial_\mu , \ \hat Z_\nu \right] &=&
2\pi i \frac{1}{L}\delta_{\mu\nu} \hat Z_\nu \ , 
\label{ncalgdelZ} 
\eeqa
where 
$\Theta_{\mu\nu} =\frac{2\pi}{L^2}\theta_{\mu\nu}$.
For the purpose that will be clear later,
we impose the commutation relation
\beq
\left[\hat\partial_\mu\,,\,\hat\partial_\nu\right]=
- i \, c_{\mu\nu} 
\label{ncalgdeldel}
\eeq
on the derivative operator $\hat\partial_\mu$.
Here the real anti-symmetric tensor
$c_{\mu\nu}$ is left arbitrary at this point,
since the derivative operator $\hat\partial_\mu$ acts 
on fields (or functions of the coordinate operators)
as adjoint, and the parameter $c_{\mu\nu}$ does 
not affect the commutativity of an action of $\hat\partial_\mu$
on to the field $\Phi$ since 
$[\hat\del_\mu,[\hat\del_\nu,\Phi]]
=[\hat\del_\nu,[\hat\del_\mu,\Phi]]$ due to the Jacobi identity.

When we impose twisted boundary conditions
on a U($p$) gauge field,
it is convenient to think of a
constant-curvature U(1) background field $\hat{A}^{(0)}_\mu$, 
which obeys the boundary conditions.
Defining the covariant derivative for the background as
\beq
\hat D^{(0)}_\mu=\hat\partial_\mu -i \hat A^{(0)}_\mu \ ,
\label{hatnabla}
\eeq
the background flux is given by
\beq
f_{\mu\nu}=i 
\left[ \hat D_\mu^{(0)}, \hat D_\nu^{(0)}\right] 
- i 
\left[\hat\partial_\mu\, , \,\hat\partial_\nu\right]
\ .
\label{noncommcurvaop}
\eeq
We decompose the gauge field into the 
background and the fluctuation as
\beq
\hat{A}_\mu = \hat{A}^{(0)}_\mu+\hat{{\cal A}}_\mu \ ,
\label{gauge-field-decomp}
\eeq
so that the boundary conditions for $\hat{{\cal A}}_\mu$
take the homogeneous form
\beq
\e^{L\hat\partial_\nu}~\hat{\cal A}_\mu~\e^{-L\hat\partial_\nu}
=\hat\Omega_\nu~\hat{\cal A}_\mu~
\hat\Omega_\nu^\dagger \ .
\label{adjointsectionop}
\eeq
The transition functions $\hat\Omega_\mu$ are chosen as
\beq
\hat\Omega_\mu 
= \e^{i \alpha_{\mu\nu} \hat x_\nu} \otimes\Gamma_\mu^{(p)} \ ,
\label{form-trans-fn}
\eeq
where $\alpha_{\mu\nu}$ are constant real values 
and $\Gamma^{(p)}_\mu$ are constant SU($p$) matrices.
The fact that the background field $\hat A^{(0)}_\mu$
obeys the boundary conditions fixes $\alpha_{\mu\nu}$.
The so-called co-cycle condition (\ref{cocycle}), 
which represents the consistency
of the boundary conditions, requires 
$\Gamma^{(p)}_\mu$ to satisfy the 't Hooft-Weyl algebra.
In the 2d case, for instance, it is given as
\beq
\Gamma^{(p)}_\mu \Gamma^{(p)}_\nu
= e^{-2\pi i \varepsilon_{\mu\nu}\frac{q}{p}}
\Gamma^{(p)}_\nu \Gamma^{(p)}_\mu \  ,
\label{pqthooftweylalg}
\eeq
where 
$\varepsilon_{\mu\nu}$ is an anti-symmetric tensor
with $\varepsilon_{12} = 1$.
Then the co-cycle condition (\ref{cocycle}) further requires
the background abelian flux $f_{12}$ to be 
given by\footnote{Although the integer $q$ 
is originally defined by (\ref{pqthooftweylalg}) 
modulo $p$,
we actually redefine it through (\ref{relfluxfq})
or (\ref{relfluxqF})
so that the theory with $q$ and that with $q+p$
are considered to be inequivalent.}
\beq
f_{12} \equiv  F = \frac{2\pi q}{L^2(p - \Theta q)} \ ,
\label{relfluxfq}
\eeq
where $\Theta_{\mu\nu} = \Theta \varepsilon_{\mu\nu}$.
Solving this for the integer $q$, we obtain
\beq
q= \frac{1}{2\pi} \frac{FL^2}{1 + \theta F} p \ ,
\label{relfluxqF}
\eeq
where $\theta_{\mu\nu} = \theta \varepsilon_{\mu\nu}$.

The surprising fact about NC geometry is that
the above gauge theory can be mapped to a 
gauge theory on a dual NC torus with periodic boundary 
conditions. This 
can be demonstrated by showing that the general 
solution to (\ref{adjointsectionop}) is given, for instance, in 2d
by (See section \ref{sec:dualtheory}.)
\beq
\hat{\cal A}_\mu
=\sum_{\vec m\in\Z^2}\,\prod_{\nu=1}^2
\left(\hat Z'_\nu\right)^{m_\nu}
\,\prod_{\lambda<\rho}\e^{-\pi i\,m_\lambda\,\Theta'_{\lambda\rho}\,m_\rho}
\, \cdot  \Bigl( \id \otimes a_\mu(\vec m) \Bigr) \ ,
\label{generaltwistsol}
\eeq
where $a_\mu(\vec m)$ are 
$p_0\times p_0$ matrices
with 
$p_0$ being the greatest common divisor of 
$p$ and $q$.
Let us introduce integers $\tilde p$ and $\tilde q$ by
\beq
p= p_0\, \tilde p \ , \ \ q= p_0\, \tilde q \ .
\label{pp0ptildeqq0qtilde}
\eeq
Since $\tilde p$ and $\tilde q$  are
co-prime, the integers $a$ and $b$ in
the Diophantine equation
\beq
a \tilde p+b \tilde q = 1 
\label{aibidef}
\eeq
are uniquely determined
up to the shift $(a,b) \sim (a,b) + (\tilde q, -\tilde p)$.
The operators $\hat{Z}'_\mu$ in (\ref{generaltwistsol})
are written in the form
\beq
\hat Z'_\mu=
\e^{2\pi i\beta_{\nu\mu}\hat x_\nu}\otimes\prod_{\nu=1}^2
(\Gamma^{(p)}_\nu)^{b\varepsilon_{\nu\mu}} \ ,
\label{hatZprime}
\eeq
where $\beta_{\nu\mu}$ is given by (\ref{defbeta1122}),
%
and $b$ is the integer appearing in (\ref{aibidef}).
One can also show that the operators
$\hat Z'_\mu$ and $\hat D^{(0)}_\mu$
satisfy the commutation relations
\beqa
\hat Z'_\mu\,\hat Z'_\nu&=&
\e^{-2\pi i\,\Theta'\varepsilon_{\mu\nu}}\,
\hat Z'_\nu\,\hat Z'_\mu \ ,
\label{Zprimealg}\\
\left[\hat D^{(0)}_\mu\,,\,\hat Z'_\nu\right]
&=&2\pi i\,
L'^{-1} \delta_{\mu\nu}\,\hat Z'_\nu \ ,
\label{nablaZprime} \\
\left[ \hat D_\mu^{(0)}, \hat D_\nu^{(0)}\right] 
&=& - i (c_{\mu\nu} + f_{\mu\nu}) \ ,
\label{nabla-nabla}
\eeqa
where the parameters $\Theta'$ and $L'$
are given by 
\beqa
\Theta'&=&\frac{a \Theta + b}{\tilde p -  \tilde q \Theta}\,
\label{Thetaprime}
\ , \\
L'&=&L \,(\tilde p - \Theta \tilde q) \ .
\label{Sigmaprime}
\eeqa

The algebra (\ref{Zprimealg})-(\ref{nabla-nabla})
have the same form as (\ref{ncalgZZ})-(\ref{ncalgdeldel})
%
for the periodic boundary conditions.
This implies that the U($p$) gauge theory with twisted boundary 
conditions on a NC torus characterized by $\Theta$ and $L$
can be mapped to a dual U($p_0$) gauge theory with periodic boundary
conditions on a NC torus characterized by $\Theta'$ and $L'$.
The covariant derivative operator $\hat D^{(0)}_\mu$
on the original torus
plays the role of the 
derivative operator $\hat\del'_\mu$ on the dual torus.
This equivalence of the two NC theories is known as
the Morita equivalence.

When the theory includes only fields in the
adjoint representation, which obey the boundary conditions
\beq
\e^{L\hat\partial_\nu}~\hat{\Phi} ~ \e^{-L\hat\partial_\nu}
=\hat\Omega_\nu~\hat{\Phi}~\hat\Omega_\nu^\dag \ ,
\label{adjsectionop}
\eeq
one can map them to fields in the dual theory with
periodic boundary conditions as we did above for the gauge field. 
Using a map from finite matrices to
fields on a discretized NC torus 
with periodic boundary conditions, one can \emph{indirectly}
regularize the
original theory with twisted boundary conditions \cite{AMNS}.
However, the Morita equivalence does not hold in general 
for theories including matter fields in the fundamental 
representation,\footnote{The particular
Morita equivalence involving fundamental matters
discussed in the second and third papers of ref.\ \cite{AMNS}
is of no use for the present purpose, since
it maps NC gauge theory with periodic boundary conditions
to a commutative gauge theory with twisted boundary conditions.}
which obey the boundary conditions
\beq
\e^{L\hat\partial_\nu}~\hat{\Phi} ~ \e^{-L\hat\partial_\nu}
=\hat\Omega_\nu~\hat{\Phi} \ .
\label{fundsectionop}
\eeq
These conditions can be solved explicitly \cite{Szabo:2001kg}, 
but the obtained solution
does not suggest any obvious way to regularize the theory
unlike the situation with the adjoint matters.
Our idea is therefore to construct the configuration space 
of a regularized field in a purely algebraic way.


\section{Rewriting boundary conditions in a covariant form}
\label{sec:solvetbc}

The gauge invariance of NC gauge theories
is represented by SU($N$) symmetry in the finite-$N$ 
matrix formulation.
Therefore our important first step is
to rewrite the boundary conditions (\ref{adjsectionop})
and (\ref{fundsectionop})
in a gauge-covariant form.

In this section, by gauge covariance we mean the covariance
under the transformation of the gauge field (\ref{gauge-field-decomp})
\beq
\hat A_\mu \to \hat g \, \hat A_\mu \, \hat g^\dag
+i \, \hat g \, [\hat\del_\mu,\, \hat g^\dag] \ ,
\label{gaugetr}
\eeq
together with the same one for the background field
\beq
\hat A_\mu ^{(0)} \to \hat g \, \hat A_\mu ^{(0)} \, \hat g^\dag
+i \, \hat g \, [\hat\del_\mu,\, \hat g^\dag]  \ ,
\label{gaugetr_bg}
\eeq
so that the covariant derivative operator $\hat D^{(0)}_\mu$
given by (\ref{hatnabla})
and the fluctuation part $\hat{\cal A}_\mu$
transform covariantly as
\beqa
\hat D^{(0)}_\mu &\to&
\hat g \, \hat D^{(0)}_\mu \, \hat g^\dag \ ,  \nonumber \\
\hat{\cal A}_\mu &\to& 
\hat g \, \hat{\cal A}_\mu \, \hat g^\dag \ .
\label{D0-calA-cov}
\eeqa
This motivates us to rewrite the twisted boundary 
conditions (\ref{adjointsectionop}) as
\beq
e^{L \hat D^{(0)}_\nu} \, 
\hat{\cal A}_\mu \, e^{-L \hat D^{(0)}_\nu}
=\hat\Xi_\nu \,  \hat{\cal A}_\mu  \, 
\hat\Xi_\nu^\dag \ ,
\label{tbccovariantform}
\eeq
where we have defined the operator
\beq
\hat\Xi_\mu
= e^{L \hat D^{(0)}_\mu} 
e^{- L \hat{\del}_\mu }
\hat\Omega_\mu \ ,
\label{Xi-def}
\eeq
which transforms covariantly as
\beq
\hat\Xi_\mu  \to \hat g \, \hat\Xi_\mu
 \, \hat g^\dag \ .
\eeq

The key observation for our formulation is that actually
$\hat\Xi_\mu$ can be written in terms of 
the coordinate operators $\hat{Z}_\mu '$ of the dual torus
that appear in (\ref{generaltwistsol}).
In 2d, for instance, 
$\hat\Xi_\mu$ and $\hat{Z}_\mu '$  are given explicitly 
as (\ref{Omega2Z1-q}) and (\ref{hatZprime2}).
Using (\ref{relfluxqF}) and (\ref{aibidef}),
one can easily show that
\beq
\hat\Xi_1
=  (\hat Z'_2)^{\tilde{q}} \quad , \quad
\hat\Xi_2
= (\hat Z'_1)^{-\tilde{q}} \ .
\label{Omega1Z2q}
\eeq
Although the above relation was obtained in
the specific gauge (\ref{backgroundgfasymgauge}),
it should hold gauge independently since both $\hat\Xi_\mu$ and
$\hat Z'_\mu$ transform covariantly.

To appreciate the meaning of (\ref{Omega1Z2q}),
it is instructive to check that
the twisted boundary conditions
(\ref{tbccovariantform}) 
are indeed satisfied 
for the solution (\ref{generaltwistsol}).
For that, it suffices to show that
\beq
e^{L \hat D^{(0)}_\nu} \, 
\hat{Z}'_\mu \, 
 e^{-L \hat D^{(0)}_\nu}
=\hat\Xi_\nu  \, \hat{Z}'_\mu \, \hat{\Xi}_\nu^\dag \ .
\label{Zprime-cond}
\eeq
Using (\ref{Zprimealg}), (\ref{nablaZprime}) and (\ref{Omega1Z2q}),
each side of eq.\ (\ref{Zprime-cond}) is given by
\beqa
e^{L \hat D^{(0)}_\mu} \ \hat Z'_\nu \ e^{-L \hat D^{(0)}_\mu}
&=& e^{2\pi i \frac{L}{L'} \delta_{\mu \nu}} \ \hat Z'_\nu \ , 
\label{eLDZpeLDZp}\\
\hat\Xi_\mu \ \hat Z'_\nu \
\hat\Xi_\mu ^{\dagger} &=& 
e^{2\pi i \Theta' \tilde{q} \delta_{\mu\nu}} \ \hat Z'_\nu \ .
\eeqa
From (\ref{Thetaprime}) and (\ref{Sigmaprime}), we obtain
\beq
\Theta' \tilde{q} = 
\frac{(a \Theta + b)\tilde{q}}{\tilde p -  \tilde q \Theta}
= 
\frac{1-a(\tilde{p}-\Theta\tilde{q})}{\tilde p -  \tilde q \Theta}
= 
\frac{1}{\tilde p -  \tilde q \Theta} - a
= 
\frac{L}{L'} -  a  \ ,
\eeq
where we have used (\ref{aibidef}). 
Hence the claim (\ref{Zprime-cond}).

The twisted boundary conditions 
(\ref{adjsectionop}) for the
adjoint matter $\Phi$, which transforms as
\beq
\hat \Phi \to \hat g \, \hat \Phi \, \hat g ^\dag  \ ,
\eeq
can be rewritten in a covariant form as
\beq
e^{L \hat D^{(0)}_\mu} \, \Phi \, e^{-L \hat D^{(0)}_\mu}
=\hat\Xi_\mu \, \Phi \, \hat\Xi_\mu^\dag \ .
\label{twistedbcadj}
\eeq
Similarly the twisted boundary conditions 
(\ref{fundsectionop}) for the
fundamental matter $\Phi$, which transforms as
\beq
\hat \Phi \to \hat g \, \hat \Phi \ ,
\eeq
can be rewritten in a covariant form as
\beq
e^{L \hat D^{(0)}_\mu} \, \Phi \, e^{-L \hat\del_\mu}
=\hat\Xi_\mu \Phi \ .
\label{twistedbcfund}
\eeq

\section{Algebraic characterization of the configuration space}
\label{sec:mmformulation}


In this section we characterize the configuration space 
of a regularized field in an algebraic way.
Here the covariant form of the twisted boundary conditions
obtained in the previous section plays a crucial role.

Let us first consider a gauge-singlet field $\Phi$, 
for which the twisted boundary conditions reduce
to the periodic ones
\beq
e^{L\hat\del_\mu} \Phi e^{-L\hat\del_\mu} = \Phi \ .
\label{perbc}
\eeq
%
Instead of considering the derivative operator
$\hat\del_\mu$, we consider only a shift operator
\beq
\hat\Gamma_\mu = e^{\epsilon \hat\del_\mu} \ ,
\eeq
where $\epsilon$ serves as the lattice spacing.
The algebra (\ref{ncalgdelZ}) and (\ref{ncalgdeldel}) 
are replaced by
\beqa
\hat\Gamma_\mu \hat Z_\nu \hat\Gamma_\mu^\dagger
&=&e^{\frac{2\pi i}{N} \delta_{\mu\nu}} \hat Z_\nu \ , 
\label{algGZGZ} \\
\hat\Gamma_\mu \hat\Gamma_\nu &=& 
e^{-i \epsilon ^2 c_{\mu\nu}}
\hat\Gamma_\nu \hat\Gamma_\mu \ ,
\label{algGGGG}
\eeqa
where the size of the torus is given by 
\beq
L=\epsilon \, N \ .
\label{size-original}
\eeq
The 
boundary conditions (\ref{perbc}) can be written as
\beq
(\hat\Gamma_\mu)^N \, \Phi \,
(\hat\Gamma_\mu^\dagger)^N  = \Phi \ .
\label{Gamma-bc-reg}
\eeq
The crucial observation here is the following.
Suppose $\Phi$ satisfies the boundary conditions (\ref{Gamma-bc-reg}).
Then so do $\hat Z_\mu \Phi$ and $\Phi \hat Z_\mu$,
as one can show easily by using the algebra (\ref{algGZGZ}).
Similarly, one finds from (\ref{algGGGG}) that
$\hat \Gamma_\mu \Phi$ and $\Phi \hat \Gamma_\mu$
obey the same boundary conditions if and only if
\beq
c_{\mu\nu} = \frac{2\pi}{N \epsilon ^2 } 
\times \mbox{integer} \ ,
\label{cond-c}
\eeq
which we shall assume in what follows.
Thus the configuration space 
can be defined as a representation space of
the operators $\hat Z_\mu$ and $\hat\Gamma_\mu$,
on which 
\beq
(\hat\Gamma_\mu)^N = \id
\label{GammaNeq1}
\eeq
is satisfied.

Let us move on to the case of adjoint matters,
which obey the twisted boundary conditions (\ref{twistedbcadj}).
The coordinate and derivative operators 
$\hat Z'_\mu$ and $\hat D^{(0)}_\mu$
on the dual torus
satisfy the algebra (\ref{Zprimealg}), (\ref{nablaZprime})
and (\ref{nabla-nabla}).
%
The regularized version of the algebra
can be constructed as follows.
Instead of the covariant derivative
$\hat D^{(0)}_\mu$, we consider only the covariant shift operator
\beq
\hat\Gamma'_\mu = e^{\epsilon \hat D^{(0)}_\mu} \ .
\eeq
The algebra (\ref{nablaZprime}) and (\ref{nabla-nabla})
should be replaced by
\beqa
\hat\Gamma'_\mu \hat Z'_\nu \hat\Gamma_\mu^{\prime\dagger}
&=&e^{\frac{2\pi i}{n} \delta_{\mu\nu}} \hat Z'_\nu \ , 
\label{algGpZpGpZp} \\
\hat\Gamma'_\mu \hat\Gamma'_\nu 
&=&e^{-i \epsilon^2 (c_{\mu\nu}+f_{\mu\nu})} 
\hat\Gamma'_\nu \hat\Gamma'_\mu \ ,
\label{algGpGpGpGp}
\eeqa
where the size of the dual torus is given by 
\beq
L'=\epsilon \, n \ .
\label{size-dual}
\eeq
The twisted boundary conditions (\ref{twistedbcadj})
can be written as
\beq
(\hat\Gamma'_\mu)^N \, \Phi \, 
(\hat\Gamma_\mu^{\prime\dagger})^N  =
 \hat\Xi_\mu \, \Phi \, 
 \hat\Xi_\mu^\dag \ .
\label{Z-bc-reg}
\eeq
From (\ref{Zprime-cond}) one finds that
\beq
(\hat\Gamma'_\mu)^N \, \hat Z'_\nu \, 
(\hat\Gamma_\mu^{\prime\dagger})^N  =
 \hat\Xi_\mu \, \hat Z'_\nu \, 
 \hat\Xi_\mu^\dag \ .
\label{GprimeZprime}
\eeq
Suppose $\Phi$ satisfies the boundary conditions (\ref{Z-bc-reg}).
Then so do $\hat Z_\mu ' \Phi$ and $\Phi \hat Z_\mu '$,
due to (\ref{GprimeZprime}).
Similarly $\hat \Gamma ' _\mu \Phi$ and $\Phi \hat \Gamma  ' _\mu$
obey the same boundary conditions if and only if (\ref{cond-c})
is satisfied.
This can be shown, for instance, in 2d by using
\beqa
(\hat\Gamma'_\mu)^N \, \hat\Gamma'_\nu
(\hat\Gamma^{' \dag}_\mu)^{N}
&=& e ^{- 2 \pi i \frac{\tilde{q}}{n} \varepsilon_{\mu\nu}
- i N \epsilon^2 c_{\mu\nu}  } 
\hat\Gamma'_\nu
 \ , \\
\hat\Xi_\mu \, \hat\Gamma'_\nu \, \hat\Xi_\mu^{\dagger}
&=& e ^{- 2 \pi i \frac{\tilde{q}}{n}  \varepsilon_{\mu\nu} }
\hat\Gamma'_\nu \ ,
\label{OmegaDOmegaDmc2}
\eeqa
which are obtained from 
(\ref{algGpZpGpZp}), (\ref{algGpGpGpGp})
and (\ref{Omega1Z2q}).
Therefore, the configuration space 
can be viewed as a representation space of
the operators $\hat Z'_\mu$ and $\hat\Gamma'_\mu$,
on which 
\beq
(\hat\Gamma'_\mu)^N = \hat\Xi_\mu 
\label{GammaNeqOmega}
\eeq
is satisfied,
where 
$\hat\Xi_\mu$ is written in terms of $\hat Z'_\mu$
as in (\ref{Omega1Z2q}).


The twisted boundary conditions on a field $\Phi$ 
in the fundamental representation are
written as (\ref{twistedbcfund}).
The regularized version is given by
\beq
(\hat\Gamma'_\mu)^N \, \Phi \, 
(\hat\Gamma_\mu^{\dagger})^N  =
 \hat\Xi_\mu \, \Phi   \ .
\label{Z-bc-reg-fun}
\eeq
Suppose $\Phi$ satisfies the boundary conditions (\ref{Z-bc-reg-fun}).
Then $\hat Z_\mu ' \Phi$ and $\Phi \hat Z_\mu $ do so,
as one can show easily by using the algebra (\ref{algGZGZ}) 
and (\ref{GprimeZprime}).
Similarly $\hat \Gamma ' _\mu \Phi$ and $\Phi \hat \Gamma _\mu$
obey the same boundary conditions if and only if (\ref{cond-c})
is satisfied.
Therefore the space of regularized configurations 
can be defined as a representation space
of the operators $\hat Z'_\mu$ and 
$\hat\Gamma'_\mu$ acting from the left,
and the operators $\hat Z_\mu$ and 
$\hat\Gamma_\mu$ acting from the right.
On the representation space,
(\ref{GammaNeq1}) and (\ref{GammaNeqOmega}) should be also satisfied.

\section{Finite-matrix formulation for twisted boundary conditions}
\label{sec:constcoodshiftope}

In this section we construct the configuration space of NC fields
explicitly as a representation space of the algebra of
coordinate and shift operators with the desired
properties discussed in the previous section.
Thus we arrive at a finite-matrix formulation,
which enables us to describe twisted boundary conditions
in terms of finite matrices.
%
Here we consider the 2d case,
but generalization to any even dimensions is straightforward.
%

First let us consider the gauge-singlet field
obeying periodic boundary conditions.
Since the 2d torus is now discretized into
a $N \times N$ lattice, it is natural to
represent a gauge-singlet field
by a $N \times N$ matrix from the counting of degrees of freedom.
Then the operators $\hat Z_\mu$, $\hat\Gamma_\mu$,
which act on it and obey 
the algebra (\ref{ncalgZZ}), (\ref{algGZGZ}) and (\ref{algGGGG}),
can be represented 
in terms of $N \times N $ matrices as
\beqa
\hat Z_\mu &=& Z_\mu^{(N)}  \ , 
\label{Zhat-repr}
\\
\hat\Gamma_\mu &=& \Gamma_\mu^{(N)} \ ,
\label{Gam-repr}
\eeqa
where $Z_\mu^{(N)}$ and $\Gamma_\mu^{(N)}$
are SU($N$) matrices satisfying the algebra
\beqa
Z_1^{(N)}  Z_2^{(N)} &=&
e^{-2 \pi i \frac{2r}{N}}  Z_2^{(N)}  Z_1^{(N)} \ .
\label{algZNN}\\
\Gamma_\mu^{(N)} Z_\nu^{(N)} \Gamma_\mu^{(N)\dag} &=& 
e^{\frac{2\pi i}{N} \delta_{\mu\nu}} Z_\nu^{(N)} \ ,
\label{algGamZNN}\\
\Gamma_1^{(N)} \Gamma_2^{(N)} &=& 
e^{2 \pi i \frac{s}{N}} \Gamma_2^{(N)} \Gamma_1^{(N)} \ .
\label{algGammaNNGammaNN}
\eeqa
The integers\footnote{The integer $2r$ needs to be
an even number for the consistency of the NC algebra
of discretized coordinates; see eq.\ (4.11) of the 
third paper of ref.\ \cite{AMNS}.
This requires $N$ to be odd since $2r$ and $N$ are co-prime.
For the same reason, 
one should
choose the integer $j$ to be even, and 
hence the integer $n$ to be odd.
Such restriction can be understood also from 
the discretized version of (\ref{Deltaprime}) 
by requiring the $\Theta '$-dependent
phase factor should have the periodicity
under shifting $m_\mu$ by units of $n$.
}
 $2r$ and $s$ are both taken to be
co-prime to $N$, which ensures the uniqueness of the
representation up to the symmetry of the algebra \cite{twistirrep}.
An explicit representation can be given in terms of 
shift and clock matrices $V_N$ and $W_N$ 
defined by (\ref{VNWNdef}). For instance,
\beqa
Z_1^{(N)} = W_N  \quad &,& \quad  Z_2^{(N)}= (V_N)^{2r} \ , 
\nonumber \\
\Gamma_1^{(N)} = V_N \quad &,& \quad \Gamma_2^{(N)} = (W_N)^s
\label{ZN-rep}
\eeqa
satisfy all the equations except
the $\mu=\nu=2$ case of (\ref{algGamZNN}),
which requires additionally the Diophantine equation
\beq
2rs-kN=-1 
\label{diophantinerskN}
\eeq
to be satisfied for some integer $k$.
Since $2r$ and $N$ are co-prime, (\ref{diophantinerskN})
fixes the integer $s$ modulo $N$.
By comparing (\ref{algZNN}) with (\ref{ncalgZZ}),
we can identify the NC parameter of the original torus 
as
\beq
\Theta = \frac{2r}{N} \ ,
\label{ncptheta}
\eeq
whereas the size of the torus is given by (\ref{size-original}).
By comparing 
(\ref{algGammaNNGammaNN}) with 
(\ref{algGGGG}), we obtain
\beq
c_{12}= -  \frac{2 \pi}{N \epsilon^2} s \ .
\label{c12}
\eeq
Therefore, the condition (\ref{cond-c}) is indeed satisfied.
Note also that the requirement (\ref{GammaNeq1}) is 
trivially satisfied.

Let us recall that in the continuum,
the parameter $c_{\mu\nu}$ in (\ref{ncalgdeldel})
is completely irrelevant and it can be left arbitrary.
However, in the regularized theory,
we need to set it to a specific non-zero value (\ref{c12}).
This is not so surprising, though, since the regularized
theory is usually more restrictive than the continuum theory. 
%
%

Next we consider the adjoint matter field
obeying twisted
boundary conditions in the U($p$) gauge theory.
Let us note that it can be mapped to
a periodic field in the U($p_0$) gauge theory
on the dual torus, which is discretized into
a $n \times n$ lattice.
Therefore, it is natural to represent the adjoint field
in the original theory by a $n p_0\times n p_0$ matrix 
from the counting of degrees of freedom.
Then the operators $\hat Z ' _\mu$, $\hat\Gamma ' _\mu$, 
which act on it and obey 
the algebra 
(\ref{Zprimealg}), 
(\ref{algGpZpGpZp}) and (\ref{algGpGpGpGp}),
can be represented 
in terms of $n p_0\times n p_0$ matrices as
\beqa
\hat Z ' _\mu &=& Z_\mu^{(n)} \otimes \id_{p_0}  \ , \\
\hat\Gamma ' _\mu &=& \Gamma_\mu^{(n)} \otimes \id_{p_0}  \ ,
\eeqa
where 
$Z_\mu^{(n)}$ and $\Gamma_\mu^{(n)}$
are SU($n$) matrices satisfying the algebra
\beqa
Z_1^{(n)}  Z_2^{(n)} &=&
e^{-2 \pi i \frac{j}{n}}  Z_2^{(n)}  Z_1^{(n)} \ ,
\label{algZn}\\
\Gamma_\mu^{(n)} Z_\nu^{(n)} \Gamma_\mu^{(n)\dag} &=& 
e^{\frac{2\pi i}{n} \delta_{\mu\nu}} Z_\nu^{(n)} \ ,
\label{algGamZn}\\
\Gamma_1^{(n)} \Gamma_2^{(n)} &=& 
e^{-2 \pi i \frac{m}{n}} \Gamma_2^{(n)} \Gamma_1^{(n)} \ . 
\label{Gammamun}
\eeqa
The integers $j$ and $m$ are both taken to be
co-prime to $n$, which ensures the uniqueness of the
representation up to the symmetry of the algebra \cite{twistirrep}.
An explicit representation can be given,
for instance, as
\beqa
Z_1^{(n)} = W_n  \quad &,& \quad  Z_2^{(n)}= (V_n)^{j} \ , 
\nonumber\\
\Gamma_1^{(n)} = V_n \quad &,& \quad \Gamma_2^{(n)} = (W_n)^{-m} \ .
\label{Zn-rep}
\eeqa
They satisfy all the equations except
the $\mu=\nu=2$ case of (\ref{algGamZn}),
which requires additionally the Diophantine equation
\beq
mj+nk'=1 \ 
\label{diophantinemjnkp}
\eeq
to be satisfied for some integer $k'$.
By comparing (\ref{algZn}) with (\ref{Zprimealg}),
we identify the NC parameter of the dual torus as
\beq
\Theta' = \frac{j}{n} \ ,
\label{ncpthetap}
\eeq
whereas the size of the dual torus is given by (\ref{size-dual}).
By comparing
(\ref{algGammaNNGammaNN}) and (\ref{Gammamun})
with 
(\ref{algGGGG}) and (\ref{algGpGpGpGp}),
and eliminating the arbitrary parameter $c_{\mu\nu}$,
%
we obtain
\beq
-2\pi i \left( \frac{m}{n}+\frac{s}{N} \right)
= -i \epsilon^2 f_{12} \ .
\label{mnsNf12}
\eeq

In our finite-matrix formulation we still need to
identify the two integers $p$ and $q$,
which characterize
the gauge theory on the NC torus with twisted boundary conditions.
We can easily identify $q= p_0 \tilde q$ from (\ref{mnsNf12})
by using (\ref{relfluxfq}) and (\ref{Sigmaprime}) as
\beq
\tilde{q}=mN+ns \ .
\label{relation-q-mn}
\eeq
With this identification,
one can show that
the requirement (\ref{GammaNeqOmega}) 
is indeed satisfied
by using the explicit representation
(\ref{Zn-rep})
and the Diophantine equation (\ref{diophantinemjnkp}).

The identification of $p=p_0\tilde{p}$, which represents the
rank of the gauge group of the original theory, 
is more indirect
since the structure of the U($p$)
gauge group is somewhat hidden
in the finite-matrix formulation.
%
We can, however, read it off from
eq.\ (\ref{Sigmaprime}), which essentially represents
the matching of the degrees of freedom on the original
torus and those on the dual torus.
Plugging (\ref{size-original}) and (\ref{size-dual})
into (\ref{Sigmaprime}), and then using
(\ref{ncptheta}), (\ref{relation-q-mn}) and (\ref{diophantinerskN}),
we obtain
\beqa
\tilde{p} &=& \frac{n}{N} + \tilde{q} \, \Theta  
\label{nNpq}
\\
 &=& 
2rm + kn \ .
\label{relation-p-mn}
\eeqa
The equations (\ref{relation-q-mn}) and
(\ref{relation-p-mn})
can be solved for $m$ and $n$ as
\beq
m=-s\tilde{p} +k\tilde{q} \ , \ \ \ 
n=N\tilde{p}-2r\tilde{q} \ .
\label{relation-mn-pq}
\eeq
This may be used to construct 
the algebra 
(\ref{algZn})-(\ref{Gammamun}) 
for the adjoint representation,
given the one for the singlet representation
(\ref{algZNN})-(\ref{algGammaNNGammaNN})
with the input of the two integers $p$ and $q$,
since the integer $j$
can be determined from
(\ref{diophantinemjnkp}).

Finally, let us check explicitly that
the NC parameter $\Theta '$ of the dual torus
is indeed given by (\ref{Thetaprime}).
Substituting (\ref{relation-q-mn}) and
(\ref{relation-p-mn}) in (\ref{aibidef}),
we obtain
\beq
m(2ar + bN) + n(ak+bs) = 1 \ .
\eeq
Comparison with (\ref{diophantinemjnkp}) yields
\beq
j = 2ar + bN  \quad \mbox{mod $n$} \ .
\eeq
Using (\ref{ncpthetap}), (\ref{nNpq}) and (\ref{ncptheta}), 
we obtain (\ref{Thetaprime}).
Thus our finite-matrix formulation of a NC torus 
with twisted boundary conditions, 
as is obvious from its construction,
realizes the Morita equivalence
at a fully regularized level.
%

For the singlet and adjoint representations,
the regularization discussed above is actually identical to 
the one in ref.\ \cite{AMNS}
except that we have now explicitly identified the
twisted boundary conditions in terms of finite matrices. 
The real advantage of our algebraic construction is that
it allows us to describe
matter fields in the fundamental representation
obeying twisted boundary conditions.
Since the operators
$Z_\mu ' $, $\Gamma '_\mu$
act from the left and
the operators $Z_\mu$, $\Gamma_\mu$ from the right,
the fundamental matter field is
naturally represented by a $n p_0 \times N$ matrix,
which is rectangular in general.


\section{Actions for the gauge field and for fundamental matters}
\label{sec:actiongauge}

In this section we construct gauge invariant actions for the 
gauge field and the matter fields using the finite-matrix
formulation described in the previous section.
The actions look formally the same as the familiar ones
for periodic boundary conditions \cite{AMNS}.
We discuss them here in detail nevertheless, since  
the size (and also the shape in the case of fundamental matter)
of the matrices has to be chosen appropriately
for the twisted boundary conditions.

When we consider path integral over the gauge field,
we fix the background field $\hat{A}_\mu^{(0)}$
once and for all, and integrate over the fluctuation
$\hat{\cal A}_\mu$.
Therefore, when we consider the 
gauge transformation (\ref{gaugetr}) in this section, we fix 
the background field $\hat{A}_\mu^{(0)}$
instead of transforming it as (\ref{gaugetr_bg}).
As a result, $\hat D^{(0)}_\mu$ and $\hat{\cal A}_\mu$
are not separately gauge covariant as in 
(\ref{D0-calA-cov}), 
but only the full covariant derivative 
\beq
\hat D_\mu = 
\hat D^{(0)}_\mu - i \hat{\cal A}_\mu
\eeq
transforms covariantly as
\beq
\hat D_\mu \to \hat g \, \hat D_\mu \, \hat{g}^\dag \ .
\eeq
Therefore, it is natural to define an operator
\beq
V_\mu \equiv e ^{\epsilon \hat{D}_\mu} \ ,
\label{defV}
\eeq
which transforms covariantly as
$V_\mu \to \hat g V_\mu \hat{g}^\dag$,
and to represent it as a $n p_0 \times n p_0$ matrix
as we did for 
$\hat{\Gamma} ' _\mu = e^{\epsilon \hat{D}_\mu ^{(0)}}$.
The gauge-invariant action for $V_\mu$
can be given by the twisted Eguchi-Kawai model
\beq
S_{\rm TEK} = -n \beta' \, \sum_{\mu \ne \nu} 
{\cal  Z}_{\nu\mu}
\tr ~\Bigl(V_\mu\,V_\nu\,V_\mu^\dag\,V_\nu^\dag\Bigr) 
 + 2 \beta' n^2 p_0 \ ,
\label{TEK-action}
\eeq
where we choose the twist ${\cal  Z}_{\nu\mu}$ to be
\beq
{\cal  Z}_{12} = \exp{\left(-2\pi i \frac{m}{n}\right)} 
\label{twistZ12}
\eeq
in 2d, for instance, in order to ensure that
the minimum of the action is given by
$V_\mu = \hat{\Gamma} '  _\mu$,
which corresponds to $\hat{\cal A}_\mu =0$ in the continuum.
The constant term in (\ref{TEK-action}) is introduced to 
make the action vanish at its minimum.

If we interpret the theory (\ref{TEK-action})
as a gauge theory on the dual torus using the Morita equivalence,
one can introduce the ``link variables''
\beq
U_\mu = V_\mu \hat{\Gamma}^{' \dagger}_\mu \ ,
\eeq
and rewrite 
(\ref{TEK-action}) as
\beq
S_{\rm TEK} = - n \beta' \, \sum_{\mu \ne \nu} 
\tr ~\Bigl\{U_\mu\, (\hat{\Gamma}^{'}_\mu 
U_\nu \hat{\Gamma}^{' \dag}_\mu)\,
(\hat{\Gamma}^{'}_\nu U_\mu^\dag \hat{\Gamma}^{' \dag}_\nu)
\, U_\nu^\dag \Bigr\}  + 2 \beta' n^2 p_0 \ ,
\label{TEKactionUhatGamman}
\eeq
which may be viewed as Wilson's plaquette action
on the discretized dual torus \cite{AMNS}.
%
%
The coefficient $\beta'$ 
can be interpreted as the lattice coupling constant,
which is related to the coupling constant $g'$
in the dual theory (\ref{dualSYM}) as \cite{AMNS}
\beq
\frac{1}{4 g'^2} =  \beta' \epsilon ^{4-D} \ .
\label{rel_gprime_betaprime}
\eeq
Note that the coupling constant $g$ of the gauge theory on
the original torus with twisted boundary conditions
is related to $g'$ through (\ref{gprime}).
Therefore, if we define the ``lattice coupling constant''
$\beta$ for the original theory by
\beq
\frac{1}{4 g^2} =  \beta \epsilon ^{4-D} \ ,
\label{rel_g_beta}
\eeq
it can be written as
\beq
\beta = \frac{1}{\tilde{p}} \left(\frac{n}{N} \right)^D \beta' \ .
\label{rel_beta_betaprime}
\eeq


Next let us consider the action for the matter field in 
the fundamental representation.
For instance, a simple gauge-invariant action for a Dirac fermion
without species doublers can be given as
\beq
S_{\rm f} = - \tr \Bigl( \bar{\Psi} D_{\rm W} \Psi \Bigr) \ ,
\label{Wilson-Dirac-action}
\eeq
where the Wilson-Dirac operator $D_{\rm W}$ can be defined as
\beq
D_{\rm W}=\frac{1}{2}\sum_{\mu=1}^D
\left\{\gamma_\mu\left(\nabla_\mu^* 
+\nabla_\mu \right) - \epsilon  \nabla_\mu^* \nabla_\mu \right\} \ ,
\label{def-Wilson-Dirac}
\eeq
using the covariant forward and backward difference operators 
$\nabla_\mu$, $\nabla_\mu^*$ defined by
\beqa
\nabla_\mu \Psi&=&
\frac{1}{\epsilon}\left(V_\mu \, \Psi \, \hat{\Gamma}_\mu ^{\dag}
- \Psi  \right) \ , \n
\nabla_\mu^* \Psi &=&
\frac{1}{\epsilon}\left(\Psi - V_\mu ^{\dagger} \,
\Psi \,  \hat{\Gamma}_\mu \right)  \  .
\label{def-cov-shift}
\eeqa
Here  $V_\mu$ is the U$(n p_0)$ matrix
introduced by (\ref{defV}),
and $\hat{\Gamma}_\mu=\e^{\epsilon \hat\partial_\mu}$ 
is the shift operator represented as (\ref{Gam-repr}).
As we mentioned at the end of section \ref{sec:constcoodshiftope},
the matter field $\Psi$ in the fundamental representation 
are represented by a $n p_0 \times N$ rectangular matrix,
and it transforms under the gauge transformation as
$\Psi \to \hat{g} \Psi$.
On the other hand, the $\bar{\Psi}$ field
is in the anti-fundamental representation,
and represented by a $N \times n p_0 $ rectangular matrix.
It transforms under the gauge transformation as
$\bar{\Psi} \to  \bar{\Psi} \hat{g}^\dag $,
and hence 
the action (\ref{Wilson-Dirac-action}) is gauge invariant.
This model may be viewed as a certain generalization
of the model \cite{Das:1983pm}
proposed to describe quarks in large-$N$ QCD
using the twisted Eguchi-Kawai model.

One can also define an analog of Neuberger's overlap 
Dirac operator \cite{Neuberger} invented originally
in lattice gauge theory, which takes 
the form\footnote{The overlap Dirac operator was
introduced on a periodic NC torus in ref.\ \cite{Nishimura:2001dq},
and the correct form of the axial anomaly has been reproduced
in the continuum limit \cite{isonagao}.
A prescription to define an analog of the overlap Dirac operator
and its index (\ref{def_nu})
on general NC manifolds including fuzzy sphere
has been proposed in ref.\ \cite{AIN2}.
}
\beq
D=\frac{1}{\epsilon} (1 - \gamma_5 \hat{\gamma}_5) \ ,
\label{def-GW-Dirac}
\eeq
where $\gamma_5$ 
is the ordinary chirality operator
and $\hat{\gamma}_5$
is the modified one defined by
\beqa
\hat\gamma_5 &=& \frac{H}{\sqrt{H^2}} \ , \\
H &=& \gamma_5 \left(1- \epsilon D_{\rm W}\right) \ .
\label{H-def}
\eeqa
In the present case, we only have to plug our
Wilson-Dirac operator $D_{\rm W}$ defined by (\ref{def-Wilson-Dirac})
into (\ref{H-def}).
The operators $\hat{\gamma}_5$ and $\gamma_5$ are used to
define the chirality for $\Psi$ and $\bar{\Psi}$, respectively,
and the Ginsparg-Wilson relation \cite{GinspargWilson}
\beq
\gamma_5 D + D \hat\gamma_5 =0
\eeq
obeyed by $D$ guarantees the exact 
chiral symmetry \cite{Luscher}.
Thanks to the index theorem \cite{Hasenfratzindex},
one can also
classify gauge configurations into topological sectors
by using the index of $D$ defined by 
\beq
\nu = \frac{1}{2} \,  \CTr \left(\gamma_5+\hat\gamma_5 \right)
= \frac{1}{2} \, \CTr \, \hat\gamma_5 \ ,
\label{def_nu}
\eeq
where the trace $\CTr$ is taken
in the configuration space of the matter field.

\section{Monte Carlo calculation of the index}
\label{sec:numerical}

In this section
we perform Monte Carlo simulations 
of the model (\ref{TEK-action}) for $D=2$,
which represents 2d NC gauge theory with twisted
boundary conditions,
and calculate
the probability distribution of the index $\nu$
of the overlap Dirac operator for the fundamental matter
defined by eq.\ (\ref{def_nu}).
See ref.\ \cite{Aoki:2006zi} for results in the case of
periodic boundary conditions.

In order to have a twisted boundary condition,
the flux (\ref{relfluxfq}) and hence the integer $q$ 
has to be nonzero.
Then the index theorem claims that
smooth gauge configurations obeying the boundary condition
should have a nontrivial index $\nu=-q$,
where the minus sign appears due to the conventions that we have adopted.
Let us choose $q=-1$.
For simplicity, we consider U(1) gauge group $p=1$,
which also implies $p_0=1$ and hence $\tilde q = -1$,
$\tilde p = 1$.
As for the integers $r$ and $k$
appearing in eq.\ (\ref{diophantinerskN}),
we choose $r=-1$, $k=-1$ (and hence $s=\frac{N+1}{2}$)
following essentially 
the choice in previous works \cite{2dU1,Aoki:2006sb,Aoki:2006zi}.
This implies, in particular, that the 
dimensionless
noncommutativity parameter (\ref{ncptheta})
is given by $\Theta = -2/N$, where $N$ represents
the size of the torus (\ref{size-original})
in units of the lattice spacing.
Note that the size $n$ of the matrices $V_\mu$
and 
the integer $m$, which labels the twist (\ref{twistZ12})
in the action (\ref{TEK-action}),
are given by $n=N-2$ and $m=-(n+1)/2$, respectively,
due to (\ref{relation-mn-pq}).
For various $n$ and $\beta '$, we measure the index
(\ref{def_nu}) for each configuration generated by Monte Carlo
simulation, and obtain 
the probability distribution $P(\nu)$, which
is normalized by $\sum_\nu P(\nu) = 1$.

In figure \ref{distrib-N-beta} we 
plot the probability distribution $P(\nu)$
obtained for various $\beta'$ at $n=15$ (left)
and for various $n$ at $\beta' = 0.55$ (right).
Note that the chosen value of $\beta'$ in the latter plot
is above the critical point $\beta_{\rm cr} ' \equiv 1/2$
of the Gross-Witten phase transition.
We find that the distribution approaches
the Kronecker delta $\delta_{\nu 1}$
not only for increasing $\beta'$ but also for increasing $n$.
Let us recall that the continuum limit of the present model should be taken
by sending $n$ and $\beta'$ to $\infty$
simultaneously with the ratio $\beta'/n$ fixed \cite{2dU1}.
It is clear from our results that
the distribution $P(\nu)$ approaches $\delta_{\nu 1}$
very rapidly in that limit.
This demonstrates that topologically nontrivial configurations 
are indeed realized
in NC gauge theory with the twisted boundary conditions
in a way consistent with the index theorem.

\FIGURE{
    \epsfig{file=eps_%
distribution_index_ns15q1.eps,%
angle=270,width=7.4cm}
    \epsfig{file=eps_%
distribution_index_planar_q1_b0.55.eps,%
angle=270,width=7.4cm}
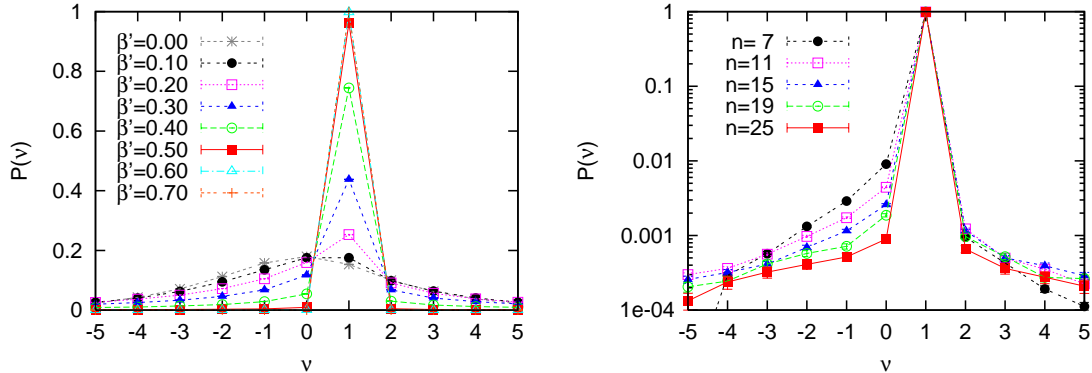
\caption{The probability distribution of the index $\nu$ 
is plotted for various $\beta '$ at $n=15$ (left) and 
for various $n$ at $\beta '=0.55$ (right).
In the latter plot, the probability is plotted in
the log scale to make the distribution at $\nu \neq 1$ visible.
}
\label{distrib-N-beta}
}

\section{Summary and discussions}
\label{sec:summary}

In this paper we have constructed a finite-matrix formulation
of gauge theories on a NC torus
in a purely algebraic way.
The configuration space has been
defined as the representation space of 
coordinate and shift operators,
which is analogous to projective modules 
in the continuum NC space.
In particular, we are able to describe
twisted boundary conditions and hence the
Morita equivalence explicitly
at a fully regularized level.
Matter fields in the fundamental 
representation are 
represented by rectangular matrices
analogously to the matrix spherical harmonics
on the fuzzy sphere with the monopole background.
By using the index of the overlap Dirac operator for the
fundamental matter, we can classify the gauge field
configurations into topological sectors.
Monte Carlo results demonstrate that topologically
nontrivial configurations survive the continuum limit
of the NC gauge theory with twisted boundary conditions
in a way consistent with the index theorem.
This also confirms the validity of our formulation
and its usefulness in various nonperturbative studies.

%


As we mentioned in the Introduction,
one of our motivations for studying topological aspects
of NC gauge theories is to understand
the realization of a chiral gauge theory in our four-dimensional
world by compactifying string theory
with a nontrivial index in the extra dimensions.
In this context, an interesting possibility would be that
NC geometry is actually realized only in the extra dimensions
as discussed in refs.\ \cite{Aschieri:2003vy} using 
the fuzzy sphere, where the dynamical generation of a nontrivial index
may be possible \cite{AIMN}.
Using our formulation, one can perform
similar analyses using a NC torus.
Considering the dramatic effects of NC geometry on
topological properties \cite{Aoki:2006zi}, 
we may hope to obtain results 
qualitatively different from what we know for commutative
extra dimensions.

Ultimately we hope to realize the whole set up
{\em dynamically},
for instance, in the IIB matrix model \cite{9612115},
which is conjectured to be a nonperturbative definition
of type IIB superstring theory in 10 dimensions.
The dynamical generation of 4d space-time in this model
is discussed, for instance in ref.\ \cite{Aoki:1998vn,gaussian}.
NC geometry appears naturally from the IIB matrix model
for particular backgrounds \cite{AIIKKT}, 
and this feature is recently focused also
in the context of emergent gravity \cite{Steinacker:2007dq}.
It would be interesting if one could describe the low energy 
effective theory of the IIB matrix model after dynamical
generation of 4d space time in terms of field theory
with NC extra dimensions.


\acknowledgments

We thank Goro Ishiki, Satoshi Iso, 
Hikaru Kawai and Asato Tsuchiya
for valuable discussions.
The work of J.N.\ is supported in part by Grant-in-Aid
for Scientific Research (Nos.\ 19340066 and 20540286) from the 
Ministry of Education, Science and Culture.


\appendix

\section{Some details of the review in Section 2}

\label{section:review}

In this Appendix we present some details of the
review given in section \ref{sec:morita} 
to make it self-contained.
We also refer the readers
to ref.\ \cite{Szabo:2001kg} for the topics that are not
covered here.

\subsection{NC gauge theory with twisted boundary conditions}

In section \ref{sec:morita} we described field theories
on a NC geometry in terms of operators.
In that language a field configuration $f(x)$ corresponds
to the 
operator $\hat f$ through
\beqa
\hat f
&=&\int d^Dx~\hat\Delta(x) f(x) \ ,
\label{mapfuncoper} \\
\hat\Delta(x)
&=& \int\frac{d^Dk}{(2\pi)^D}~
\e^{ik_\mu \hat x_\mu}~e^{-ik_\mu x^\mu} \ ,
\label{Deltadef}
\eeqa
where the coordinate operators $\hat x_\mu$ and
the derivative operators $\hat\partial_\mu$ 
satisfy the algebra (\ref{noncommalg}).
The product of two fields $f(x)$ and $g(x)$ 
are defined by the operator product of the corresponding
operators $\hat{f}$ and $\hat{g}$, and it can be
given explicitly by the so-called Moyal star-product 
\beq
f(x)\star g(x) =f(x)~\exp\left(\frac i2\,\overleftarrow{\partial_\mu}\,
\theta_{\mu\nu}\,\overrightarrow{\partial_\nu}\right)~g(x) \ ,
\label{starproddef}
\eeq
where $\theta_{\mu\nu}$ is the NC parameter appearing 
in (\ref{noncommalg}).

Let us consider a U($p$) gauge theory on a NC torus,
whose action is given by 
\beqa
S_{\rm YM} &=& \frac1{4g^2}\,\int_{{\bf T}^D} d^Dx~\tr^{~}_p
\Bigl(F_{\mu \nu}(x)-f_{\mu \nu}\, \Bigr)_\star^2 \ ,
\label{SYMmulti} \\
F_{\mu\nu}(x) &=& \partial_\mu A_\nu(x) -\partial_\nu A_\mu(x)
-i\Bigl(A_\mu(x) \star A_\nu(x)-A_\nu(x) \star A_\mu(x) \Bigr) \ .
\eeqa
The constant background flux $f_{\mu\nu}$
will be specified later.
We require the gauge field $A_\mu(x)$ 
to obey the twisted boundary conditions
\beq
A_\mu (x+L \hat{\nu})
=\Omega_\nu(x)\star A_\mu (x) \star \Omega_\nu (x)^\dagger
+i\,\Omega_\nu (x)\star\partial_\mu \,\Omega_\nu (x)^\dagger \ ,
\label{twistedbc}
\eeq
where $\Omega_\nu (x)$ are the transition functions,
which are $p\times p$ star-unitary matrices.
The symbol $\hat \nu$ represents a unit vector in the $\nu $ direction.
Consistency of the conditions (\ref{twistedbc}) requires 
the transition functions $\Omega_\mu (x)$ to satisfy the 
co-cycle conditions
\beq
\Omega_\mu(x+L\hat\nu)\star\Omega_\nu(x)=
\Omega_\nu(x+L\hat\mu)\star\Omega_\mu(x) \ .
\label{cocycle}
\eeq

It is convenient to 
introduce a background abelian gauge field $A^{(0)}_\mu(x)$,
which obeys the twisted boundary conditions (\ref{twistedbc}),
and to decompose the gauge field configuration $A_\mu(x)$ 
into the background and the fluctuation as
\beq
A_\mu(x)=A^{(0)}_\mu(x)+{\cal A}_\mu(x) \ .
\eeq
Then the boundary conditions for the fluctuation
${\cal A}_\mu(x)$ take the homogeneous form as
\beq
{\cal A}_\mu(x+L \hat{\nu})
=\Omega_\nu (x)\star{\cal A}_\mu (x)\star\Omega_\nu (x)^\dagger \ .
\label{adjointsection}
\eeq
In order for $A^{(0)}_\mu(x)$ to give the minimum of the
classical action (\ref{SYMmulti}), we choose
the flux $f_{\mu\nu}$ in (\ref{SYMmulti}) to be
\beq
f_{\mu\nu}
=\partial_\mu A^{(0)}_\nu(x)-\partial_\nu A^{(0)}_\mu(x)
-i\left(A^{(0)}_\mu(x) \star A^{(0)}_\nu(x)
-A^{(0)}_\nu(x) \star A^{(0)}_\mu(x) \right) \ .
\label{noncommcurva}
\eeq
Then the action (\ref{SYMmulti}) can be rewritten as
\beqa
S_{\rm YM} &=& \frac1{4g^2}\,\int_{{\bf T}^D} d^Dx~\tr^{~}_p
\Bigl({\cal F}_{\mu\nu}(x)\Bigr)_\star^2 \ ,
\label{SYMsingle} \\
{\cal F}_{\mu\nu}(x)
&=&D^{(0)}_\mu{\cal A}_\nu(x)-D^{(0)}_\nu{\cal A}_\mu(x)
-i\Bigl({\cal A}_\mu(x)\star{\cal A}_\nu(x)
-{\cal A}_\nu(x)\star{\cal A}_\mu(x)\Bigr) \ ,
\label{calFdef}
\eeqa
where we have defined the covariant derivative $D^{(0)}_\mu$ with the 
background field as
\beq
D^{(0)}_\mu=\partial_\mu-i \, [A^{(0)}_\mu(x) \, , ~~~ ]_\star \ .
\label{nabladef}
\eeq

Using the map (\ref{mapfuncoper}),
we can define operators $\hat A_\mu$,
$\hat A^{(0)}_\mu$, $\hat{\cal A}_\mu$ and 
$\hat\Omega_\mu$, which correspond
to the fields $A_\mu(x)$, $A^{(0)}_\mu(x)$, 
${\cal A}_\mu(x)$ and $\Omega_\mu(x)$,
respectively.
The action (\ref{SYMsingle}) can be written as
\beq
S_{\rm YM}= \frac1{4g^2}\,\Tr\otimes\tr^{~}_p
\left(\left[\hat D^{(0)}_\mu\,,\,\hat{\cal A}_\nu\right]
-\left[\hat D^{(0)}_\nu\,,\,\hat{\cal A}_\mu \right]
-i\left[\hat{\cal A}_\mu\,,\,\hat{\cal A}_\nu\right]\right)^2 \ ,
\label{SYMop}
\eeq
where $\hat D^{(0)}_\mu$ is given by (\ref{hatnabla}).
The twisted boundary conditions (\ref{adjointsection}) 
for ${\cal A}_\mu(x)$ can be written as (\ref{adjointsectionop}).
The background flux (\ref{noncommcurva}) is written as
(\ref{noncommcurvaop}).

\subsection{Explicit forms of the background gauge field 
and the transition functions}
\label{explicitA0-Omega}

Let us consider the 2d case, and take the background gauge field as
\beq
\hat A^{(0)}_1 = 0 \ , \ \
\hat A^{(0)}_2 = \hat{x}_1 F \otimes \id_p \ .
\label{backgroundgfasymgauge}
\eeq
Then the covariant derivative operators are given as
\beq
\hat D_1^{(0)} = \hat{\del}_1\otimes \id_p \ , \ \ 
\hat D_2^{(0)} = (\hat{\del}_2 -i \hat{x}_1 F)\otimes \id_p \ ,
\label{covdelbgasymgauge}
\eeq
and the background abelian flux is obtained as
$f_{12} = F$ by using (\ref{noncommcurvaop}).

We also assume that the transition function takes the form
(\ref{form-trans-fn}), where
$\alpha_{\mu\nu}$ should be determined by requiring
the background field (\ref{backgroundgfasymgauge})
to obey the twisted boundary conditions
\beq
\e^{L\hat\partial_\nu}~
\hat A^{(0)}_\mu
~\e^{-L\hat\partial_\nu}
=\hat\Omega_\nu~\hat A^{(0)}_\mu~
\hat\Omega_\nu^\dagger 
+i \, \hat\Omega_\nu \, 
[\hat{\del}_\mu , \hat\Omega_\nu^\dagger  ] \ .
\label{adjointsectionop2}
\eeq
Then the transition functions become
\beq
\hat\Omega_1 = \e^{i \frac{LF}{1 + \theta F} \hat{x}_2} 
\otimes\Gamma_1^{(p)} \ , \ \
\hat\Omega_2 = \id\otimes\Gamma_2^{(p)} \ ,
\label{transitionfuncasymgauge}
\eeq
where $\theta_{\mu\nu} = \theta \varepsilon_{\mu\nu}$.
Imposing the co-cycle conditions (\ref{cocycle}) on them,
one can easily obtain (\ref{relfluxqF}).
The covariantized transition functions
$\hat{\Xi}$ defined by (\ref{Xi-def})
can be expressed as
\beqa
\hat\Xi_1 
&=& e^{i \frac{1}{1+\theta F}\hat{x}_2 FL} \otimes \Gamma^{(p)}_1
 \ , \nonumber
\\
\hat\Xi_2 
&=& e^{-i \hat{x}_1 FL} \otimes \Gamma^{(p)}_2  \ . 
\label{Omega2Z1-q} 
\eeqa

An explicit representation for $\Gamma_\mu^{(p)}$, which satisfy
the 't Hooft-Weyl algebra (\ref{pqthooftweylalg}),
can be given as\footnote{The representation
of the 't Hooft-Weyl algebra in any even dimension
can be obtained similarly \cite{twistirrep}.}
\beqa
\Gamma^{(p)}_{1}&=& V_{\tilde p}
\otimes\id_{p_0}
 \ , \n
\Gamma^{(p)}_{2}&=&\left(W_{\tilde p}\right)^{-\tilde q} 
\otimes\id_{p_0}
 \ ,
\label{Gammadefs}
\eeqa
where $V_{\tilde p}$ and $W_{\tilde p}$ are 
the SU($\tilde p$) shift and clock matrices
\beq
V_{\tilde p}=\begin{pmatrix}
0&1& & &0\cr &0&1& & \cr & &\ddots&\ddots& \cr
 & & &\ddots&1\cr1& & & &0\cr
\end{pmatrix}  \ , \ \
W_{\tilde p}=\begin{pmatrix}
1& & & & \cr
&\e^{2\pi i/{\tilde p}}& & &\cr & &\e^{4\pi i/{\tilde p}}& & \cr
 & & &\ddots& \cr & & & &\e^{2\pi i({\tilde p}-1)/{\tilde p}}\cr
 \end{pmatrix} 
\label{VNWNdef}
\eeq
obeying the commutation relations
\beq
V_{\tilde p} W_{\tilde p}=
\e^{2\pi i/{\tilde p}}\,W_{\tilde p}V_{\tilde p} \ .
\label{VWalg}
\eeq

\subsection{Morita equivalence}
\label{sec:dualtheory}

We are now ready to solve the
twisted boundary conditions (\ref{adjointsectionop}).
For any pair of co-prime integers $\tilde p, \tilde q$,  the set
$\{(V_{\tilde p})^j\,(W_{\tilde p})^{\tilde{q}j'}~|~j,j'\in\Z_{\tilde p}\}$ 
spans the $\tilde{p}^2$ dimensional complex
linear vector space. 
We therefore expand $\hat{\cal A}_\mu$ as
\beq
\hat{\cal A}_\mu =\int d^2 k~\e^{ik_\nu\hat x_\nu}
\otimes\sum_{\vec j~{\rm mod}\,\tilde p}~\prod_{\lambda=1}^2
(\Gamma^{(\tilde p)}_\lambda)^{j_\lambda}
\otimes\tilde a_\mu(\vec k,\vec j\,) \ ,
\label{WeylcalAexp}
\eeq
where $\tilde a_\mu(\vec k,\vec j\,)$ is a $p_0\times p_0$ 
matrix-valued function
which has a periodicity $\tilde p$ in $\vec j$.
By applying the conditions (\ref{adjointsectionop})
to (\ref{WeylcalAexp}),
and using (\ref{transitionfuncasymgauge}),
we find that the functions $\tilde a_\mu(\vec k,\vec j\,)$
should vanish unless
\beq
\frac1{2\pi}\frac{1}{\tilde p}\left(\beta^{-1}\right)_{\mu\nu} k_\nu
+\frac{\tilde q}{\tilde p} \varepsilon_{\mu\nu}j_\nu=n_\mu
\label{tildeavanish}
\eeq
for some $n_\mu \in \Z$,
where $\beta_{\mu\nu}$ are defined as
\beq
\beta_{11}= \frac{1+ \theta F}{L \tilde p} \ , \ \ 
\beta_{22} = \frac{1}{L \tilde p} \ ,
\label{defbeta1122}
\eeq 
and $\beta_{\mu\nu}=0$ for $\mu \neq \nu$.
Using the integers $a,b$ introduced in (\ref{aibidef}),
we can solve\footnote{Note 
that since $\tilde p$ and $\tilde q$ are co-prime,
this solution for $(n_\mu,j_\mu)$ is unique
for any $m_\mu$ up to a shift
$({\tilde q} l_\mu, -{\tilde p} \varepsilon_{\mu\nu} l_\nu)$
with arbitrary integer vector $l_\mu$.
This shows that (\ref{generaltwistsol}) is the unique solution
for (\ref{adjointsectionop}).
}
(\ref{tildeavanish})
by setting $n_\mu=a\,m_\mu$ and
$j_\mu=b \, \varepsilon_{\mu\nu}\,m_\nu$ 
for some $m_\mu\in\Z$.
Thus we obtain the momenta as 
$k_\mu=2\pi\beta_{\mu\nu}\,m_\nu$ and replace the
integration in (\ref{WeylcalAexp}) by 
a sum over all $\vec m\in\Z^2$.

Therefore, the general solution to the conditions
(\ref{adjointsectionop}) takes the form (\ref{generaltwistsol}),
where $a_\mu(\vec m)=
\tilde a_\mu(2\pi\beta_{\mu\nu}m_\nu,b\varepsilon_{\mu\nu}m_\nu)$ are 
$p_0\times p_0$ matrix-valued Fourier coefficients.
The operators $\hat Z'_\mu$ are defined by (\ref{hatZprime}),
and more explicitly, they are given as
\beqa
\hat Z'_1
&=& e^{2\pi i \frac{1+\theta F}{L\tilde{p}} \hat{x}_1} \otimes 
(\Gamma_2 ^{(p)})^{-b}  \ ,
\nonumber \\
\hat Z'_2
&=& e^{2\pi i \frac{1}{L\tilde{p}} \hat x_2} 
\otimes (\Gamma_1 ^{(p)})^{b} \ .
\label{hatZprime2}
\eeqa
They are shown to satisfy the algebra 
(\ref{Zprimealg}) and (\ref{nablaZprime}),
and thus they are interpreted as the coordinate operators 
of the dual torus.
While the derivation has been
given in the specific gauge (\ref{backgroundgfasymgauge}),
the results (\ref{Zprimealg}) and (\ref{nablaZprime}) are gauge covariant,
and the relations (\ref{Thetaprime}) and (\ref{Sigmaprime})
are gauge invariant.\footnote{Indeed, 
the results agree with those obtained in
ref.\ \cite{AMNS,Szabo:2001kg} with
the symmetric gauge,
which is more useful for higher dimensional extensions.}

One can map the operators to periodic fields
on the dual torus by
\beqa
\label{fieldAprime}
{\cal A}'_\mu (x')
&=& \Tr' \left(\hat\Delta'(x') \hat{\cal A}_\mu \right) \\
\hat\Delta'(x') 
&=& \frac{1}{L'^2}\sum_{\vec m\in\Z^2}\,\prod_{\nu=1}^2
\left(\hat Z'_\nu\right)^{m_\nu}
\,\prod_{\lambda<\rho}\e^{-\pi i\,m_\lambda\,\Theta'_{\lambda\rho}\,m_\rho}
\e^{-2\pi i\frac{1}{L'} m_\mu x'_\mu} \ ,
\label{Deltaprime}
\eeqa
since $\hat{\cal A}_\mu$ can be expanded in terms of
$\hat Z'_\mu$
as in (\ref{generaltwistsol}).

In terms of the fields (\ref{fieldAprime}),
the action (\ref{SYMop}) can be written 
in arbitrary dimension $D$ as
\beqa
S_{\rm YM} &=& \frac1{4g'^2}\,\int d^Dx'~\tr^{~}_{p_0}
\Bigl({\cal F}_{\mu\nu}'(x') \Bigr)^2_{\star'} \ ,
\label{dualSYM}\\
{\cal F}_{\mu\nu}'(x') &=&
\partial_\mu'{\cal A}_\nu'(x')-\partial_\nu'{\cal A}_\mu'(x')
-i\Bigl({\cal A}_\mu'(x')\star'{\cal A}_\nu'(x')
-{\cal A}_\nu'(x')\star'{\cal A}_\mu'(x')\Bigr) \ ,
\label{calFprime}
\eeqa
where $\star'$ denotes the new star-product with the
NC parameter 
$\theta' _{\mu\nu}=\frac{L'^2}{2\pi} \Theta' _{\mu\nu}$ 
instead of $\theta_{\mu\nu}$,
and $\partial_\mu'$ is the ordinary derivative operator
on the dual NC torus.
Since the operator trace $\Tr'$ is 
related to the original trace $\rm{Tr}$ by
\beq
\Tr'\otimes\tr^{~}_{p_0}=\frac{p_0}{p}\,
\left|\frac{L'}{L}\right|^D\,\Tr\otimes\tr^{~}_p \ ,
\label{traceprimes}
\eeq
the dual gauge coupling constant in (\ref{dualSYM}) is 
related to the original one in (\ref{SYMsingle}) as
\beq
g'^2=g^2\, \frac{p_0}{p}\, \left| \frac{L'}{L}\right|^{D} \ .
\label{gprime}
\eeq



\end{document}